\documentclass[preprint,aps,tightenlines,showpacs]{revtex4}
\usepackage{amsmath,amsfonts,amssymb,amsthm}
\usepackage{color}
\usepackage{graphicx,latexsym}

\begin{document}

\title{Observable eigenstate overlap in a nonlinear\\ mean-field quantum model}
\author{Claude G.\ Reinisch}
\affiliation{Science Institute, University of Iceland, Dunhaga 3, IS-107 Reykjavik,
             Iceland}
\begin{abstract}
The soliton effect is defined in nonlinear physics by the transformation of
 a {\it nonlinear} time-dependent dynamical system  into an equivalent
{\it linear}   spectral eigenproblem whose invariant
 eigenvalues
unambiguoulsly define all the    dynamical    properties of the original
  system. We point out the existence of such an effect in
a non-relativistic  isotropic two-electron mean-field quantum-dot
model. It  yields the prediction of 
observable  modulation of radiation absorption
 between its  two  lowest-energy zero-angular-momentum
  nonlinear eigenstates
(i.e.\ eigenstates which   include 
electron-electron interaction: hence their non-orthogonality). Characteristic values for  such  a possible experiment
are given in the case of GaAs.
Furthermore it provides an intriguing
nonlinear definition of the fine-structure constant solely in terms of these eigenstates. 
\end{abstract}


\maketitle

\section{Introduction}

 Mean-field nonlinear quantum approaches have become fruitful  in the description
of  manybody   quantum systems 
\cite{Pang05}. The resulting nonlinear eigenstates are
quite useful in the definition of their  basic stationary properties \cite{Reinisch04:033613}
\cite{Reinisch07:120402}.
In some cases (e.g. eigenstates with  same
 angular momenta), these nonlinear eigenstates 
are not orthogonal and thus not observable 
\cite{Huttner96:54}. 
Therefore, they should rather be regarded as convenient mathematical tools (similar to trial functions in 
variational problems) whose interest lies in their accuracy to provide observable results.
Surprisingly enough (since mean-field  descriptions originally address manyparticle
quantum systems)  they also appear
  quite useful   in the physical description of   a  single couple  of  bound-state
interacting electrons;  e.g.\       highly-compressed  
astrophysical helium
\cite{Reinisch01:042505};   
Schr{\"o}dinger-Poisson      theory for quantum-dot helium \cite{Reinisch12:902}
and related nonlinear interference effects \cite{Reinisch11:699};
 composite-fermion flux quantization
 in  Fractional Quantum Hall Effect \cite{Reinisch14:378};
excitation of radial collective modes  beyond   linear response \cite{Gudmundsson14:526}.
Note that   a  $S=0$ oscillating  electron pair  
 was investigated   
by use of
  SP nonlinear eigenstates    in 
an early  attempt  to provide a pioneering description of superconductivity
prior to   BCS theory   \cite{Schafroth55:463}.  

Nevertheless,
a formal link ---if any--- between linear and nonlinear quantum eigenstates is still
missing. One possible way to establish it is by  reference to  the
otherwise well-known soliton paradigm in nonlinear physics. It
 consists in   transforming a nonlinear dynamical system  into an equivalent
linear   spectral eigenproblem 
defined by a specific {\it ad hoc}   
Hamiltonian ${\cal H}_{sol}$
whose time-independent
 eigenvalues
unambiguoulsly yield  all the  invariant   physical    properties of the original
 system
\cite{Bullough80}  \cite{Dodd82}.
The crucial step  in this 
so-called ``inverse scattering transform"  consists in defining ${\cal H}_{sol}$
 from the  structural   properties of the      nonlinear   system itself.
In the pioneering 1+1 d Korteweg-DeVries  example \cite{Remoissenet99}, 
${\cal H}_{sol}$ is simply Schr{\"o}dinger-like with a confining ---or ``soliton"--- potential 
merely  equal at each time $t$ 
to the original time-dependent 
nonlinear wave profile multiplied by $-1$. Although this soliton  potential is thus time-dependent,
its discrete eigenvalues are not: they are constant and each of them define a single stable 
(collision-proof: hence its name) asymptotic propagating soliton.
Note that the  linear  eigenfunctions  corresponding to these invariant eigenvalues are only technical ---or virtual--- 
mathematical intermediates
without any physical meaning. Exactly like the  
interacting virtual-photons eigenstates 
of ${\cal H}_{sol}$ shown further below (Section IV).

 The aim of the present work
is to point out the existence of such soliton transformation
in isotropic quantum-dot helium \cite {Pfannkuche93:6} \cite {Pfannkuche93:2244}
and to draw observable consequences from it.
The  two-state nonlinear Schr{\"o}dinger-Poisson   (SP)
quantum model is
defined and   discussed  in Section II. 
Firstly, we
introduce  two  real-valued  SP  eigenstates   $|a)$ and $|b)$ defined by their  respective  eigenvalues 
$\mu_a$ and $\mu_b$.  We use parenthesis instead of brackets   in order to outline their
  non-orthogonality 
defined by the   non-zero inner product --or eigenstate overlap-- $( a | b )\neq 0$.
Then we show in Section III  by use of the lowest-order 
time-dependent Schr{\"o}dinger propagator \cite{Feynman19:65}  
that the    transition amplitude induced by  eigenstate overlap
from    ground state $|a)$
   to   excited eigenstate $|b)$ 
is equal, up to the mere phase factor $e^{-i \mu_b t/\hbar}$,  to $( a | b )$. Therefore 
the corresponding leading-order  transition probability $\Pi= |( a | b ) e^{-i \mu_b t/\hbar}|^2$ 
from   $|a)$
  to  $|b)$
is constant and equals $( a | b )^2$. This property resembles the
above-mentioned soliton paradigm. Indeed we show that
it can be recovered by the solution  of 
a specific          linear  spectral  eigenproblem
  defined by  an  appropriate  electromagnetic Hamiltonian which we construct
in Section IV 
 by use of 
quantum electrodynamics (QED).
Specifically, it  is built up   from the    properties of the two-state nonlinear   quantum model
in a quite similar  way as the  soliton   potential ${\cal H}_{sol}$
is extracted  from the structural   properties of  the original nonlinear system.
Then, we compare in Section V  the
 quantum predictions
of these two models  ---namely, the nonlinear SP differential system
versus  its corresponding soliton-like linear eigenproblem--- within the unavoidable error bar $\sim1 \%$
due to \underbar{self-consistency}. Indeed  our   nonlinear SP quantum model makes
an explicit use of   classical
Coulomb interaction   defined by  its electrostatic Poisson equation. 
On the other hand,  only the  
lowest-order   term of  the QED perturbation series
(which amounts to a $\sim1 \%$ error
by considering  one single electron-photon scattering per particle)   corresponds to
such a  classical Coulomb potential 
  \cite{Lawrie90}. Therefore  only  this   leading  term
should be kept in order to
construct from QED the soliton Hamiltonian ${\cal H}_{sol}$.  
Then we check that the linear eigenproblem defined by
${\cal H}_{sol}$ yields indeed
 mean quantum properties that are equivalent to those of
the original SP nonlinear model
at maximum eigenstate overlap. 
In Section VI, we provide some hints for  
experimental observation of such a soliton  effect and conclude
by a summary of the main results in Section VII.
 
\section {  nonlinear  stationary two-state quantum system}

Consider
a $S=0$ couple  of   
opposite-spin electrons of particle mass $m_e$
confined in  an external
 isotropic harmonic potential  of frequency $\omega$.
Discard the center-of-mass motion which    separates out anyway,
due to the generalized Kohn theorem related to the assumption
of parabolic confinement  \cite{Reimann02:74} \cite{Kohn61:123} and 
approximate  the main properties of
the    internal structure of the two-particle wave function by use of   a
single-particle  two-state 
mean-field  nonlinear differential
     model. Actually, such a simple model    yields quite acceptable
results
\cite{Reinisch12:902}.  Select further  those two eigenstates  
---say
    ground-state  $|a)$ or      excited eigenstate  
 $|b)$)---
where the two particles  lie 
\underbar{both}  in the same  $s$   orbital
state (no angular momentum)
  in agreement with  
Pauli exclusion principle.
In this configuration there is neither exchange energy 
nor spin-orbit coupling. However  correlation effects described by Density-Functional Theory
\cite{Reimann02:74}
will not be taken into account in the present  mean-field model.
Therefore our  two-state  
quantum system  
defines
 (in CGS units) 
the \underbar{same} single-particle
internal
 eigenstate $\Psi_{i}$ (i=a,\,b)  
for both particles
by use of the following  SP
nonlinear
differential system  
\cite{Reinisch12:902}: 
\begin{equation}
\label{eq-schroe3d}
-\frac{\hbar^2}{2m_e}\Bigl[ \nabla^2  + 
\frac{1}{2} m_e\omega^2 r^2 + e\Phi_{i}
\Bigr]     \Psi_{i}=\mu_{i} \Psi_{i},
\end{equation}
\begin{equation}
\label{eq-poisson3d}
 \nabla^2 \Phi_{i}=-4\pi e \bigl|\Psi_{i}   \bigr|^2,
\end{equation}
where  $r^2=x^2+y^2+z^2$ and 
$\nabla^2={d^2}/{dr^2} +   ({2}/{r}) ({d}/{dr})$ in
 3d radial symmetry.
Besides the external confining frequency
$\omega$ ---the sole tunable parameter of  the above differential system---,
the only fundamental parameters  
are   $\hbar$ together with $e$ and $m_e$ 
(the electron's charge and mass). In particular, there is
no velocity of light $c$, due to the evidently non-relativistic
description of both the Schr{\"o}dinger wave funtion $\Psi_{i}$ by use of  Eq.~(\ref{eq-schroe3d})
and the classical Coulomb electrostatic  interaction $\Phi_{i}$
 by use of Eq.~(\ref{eq-poisson3d}).

Defining the harmonic length  $L =\sqrt{\hbar/(2m_e \omega)}$    and
\begin{equation}
\label{eq-N}
{\cal N}= e^2{\tilde /}L =  
\frac{e^2/L}{\hbar\omega}\propto \frac{1}{\sqrt{\omega}},
\end{equation}
as the characteristic     dimensionless      nonlinearity 
--or Coulomb-interaction-- parameter   of our model (depending   only on frequency $\omega$
of the external harmonic trap),   Eqs (\ref{eq-schroe3d}~-~\ref{eq-poisson3d}) can be put in dimensionless
form by using
$C_i  ={\tilde \mu}_i -e{\tilde \Phi  }_i $,  $u_i   =\sqrt{4\pi {\cal N} L^3}\,\Psi_i   $ and $X=r/L$
while
 the  tilde superscript defines energies  in units of $\hbar\omega$. Equations
 (\ref{eq-schroe3d} - \ref{eq-poisson3d} ) then become:
\begin{equation}
\label{eq-schroeReduite}
\Biggl[ \frac{d^2}{dX^2} +   \frac{2}{X}\frac{d}{dX}  + C_i    -
\frac{1}{4}X^2
\Biggr]  u_i   =0,
\end{equation}
\begin{equation}
\label{eq-poissonReduit}
\Biggl[ \frac{d^2}{dX^2} +   \frac{2}{X}\frac{d}{dX}\Biggr] C_i   =u^2_i  .
\end{equation}
We check that the  non-interacting linear limit is 
 obtained for ${\cal N}\rightarrow 0$ which means 
$\omega\rightarrow \infty$
  and
  $u^2\propto {\cal N} \sim 0$ in the r.h.s. of Eq.~(\ref{eq-poissonReduit}).
 Then, Eqs.~(\ref{eq-schroeReduite}) and (\ref{eq-poissonReduit})
 become uncoupled and Poisson Eq.\  (\ref{eq-poissonReduit}) yields  
  vanishing particle-particle interaction $\Phi_i   \sim 0$
while (Eq.\ \ref{eq-schroeReduite})  defines   the standard 3d 
one-particle  linear isotropic harmonic-oscillator eigenstate.

The single-particle normalization $4\pi\int_0^{\infty}|\Psi_i   |^2 r^2 dr = 1$
yields ($i=a,\,b$):
\begin{equation}
\label{eq-normalisation}
 \int_0^{\infty}u ^2_i  X^2 dX = {\cal N}.
\end{equation}
For a given value of nonlinearity parameter ${\cal N}$,  the eigenstate overlap    between 
 $|a)$ and 
 $|b)$
becomes in these dimensionless variables:
\begin{equation}
\label{eq-ProdScal}
( a|b )=\frac{1}{{\cal N}} \int_0^{\infty}u_{a}u_{b} X^2 dX ,
\end{equation}
where:
\begin{equation}
\label{eq-EqualN}
\int_0^{\infty}u_a ^2 X^2 dX =\int_0^{\infty}u_b ^2 X^2 dX={\cal N},
\end{equation}
in accordance with Eq.~ (\ref{eq-normalisation})  (indeed the two eigenstates are 
located in the same harmonic trap
defined by $\omega$).
Given  $\omega$    and hence   ${\cal N}$ from  
Eq.~(\ref{eq-N}), 
the integro-differential system of coupled equations (\ref{eq-schroeReduite}-\ref{eq-EqualN}) is numerically
solved by using the 
appropriate  ``no-cusp''    initial conditions $u_{a,b}  (0)\sim\sqrt{{\cal N}}$ (given), 
$[du_{a,b}  /dX]_{X=0}=0$ and 
$C_{a,b}  (0)$ (given), $[dC_{a,b}  /dX]_{X=0}=0$ in order to select, under equal-norm condition  Eq.~(\ref{eq-EqualN}),    
   the eigensolutions $u_{a,b}  (X)$ defined by their respective regular 
boundary conditions: $\lim_{X\rightarrow\infty}u_{a,b}  (X)=0$ 
and    corresponding coupled potential
solutions
 $C_{a,b}  (X)$.  Then,   
eigenstate overlap $(a|b )$ is  obtained by Eq.~ (\ref{eq-ProdScal}).  It
also reads  by use of Hermiticity of the Laplacian operator \cite{Bec10:00}:
\begin{equation}
\label{eq-theoremW}
(a|b)= \frac{W_{ab}}{\mu_b-\mu_a},
\end{equation}
where 
$W_{ab}=( a|W|b)$ is the matrix element of Coulomb potential
\begin{equation}
\label{eq-W}
W=e(\Phi_{b}-\Phi_{a}),
\end{equation}
related to eigenstates $|a, b)$ and  to
  their respective eigenvalues $\mu_{a,b}$.
Single-particle
Coulomb-interaction potentials $\Phi_{a,b}$ are defined
in accordance with 
 Eqs.~(\ref{eq-schroe3d} - \ref{eq-poisson3d}).
In the above dimensionless  variables, 
the eigenvalue ${\tilde \mu}_i=C_i(X) + e {\tilde \Phi}_i(X)$  ($i= a,\,  b$)   can be calculated  
by use of either the   initial  conditions or  the  boundary conditions (resp. $X=0$ or $X\rightarrow\infty$: this latter being 
analytically derived from Poisson equation (\ref{eq-poissonReduit}) ):
\begin{equation}
\label{eq-mu}
 {\tilde \mu}_i=C_i(0) + e {\tilde \Phi}_i(0)=\lim_{X\rightarrow\infty}\Bigl[ C_i(X) + \frac{{\cal N}}{X} \Bigr],
\end{equation}
where 
\begin{equation}
\label{eq-greenfct}
 e{\tilde \Phi}_i(0)=\int_0^{\infty} G(0,X) u^2_i  X^2dX=\int_0^{\infty} \frac{1}{X}u^2_i  X^2dX ,
= \int_0^{\infty}  u^2_i  X dX,
\end{equation}
due to 
 3d Green function $G(X',X) = 1/|X'-X|$ 
of   Eq.\ (\ref{eq-poissonReduit}) at $X'=0$. Equations  (\ref{eq-mu} - \ref{eq-greenfct}) provide an excellent
test for the accuracy  of the numerical code   (we actually obtained a $10^{-8}$ precision by use of 
MatLab's ode45 integration code).

\section { time-dependent description of   nonlinearly-induced  transitions}

Let us first
describe   nonlinearly-induced (or ``intrinsic")  transitions
between both  eigenstates $|a,b)$ of 
the above
system due to eigenstate overlap defined   by Eqs.~(\ref{eq-theoremW} - \ref{eq-W}).  
Assume  that
the  system whose time-dependent
wavefunction is $\Psi(r,t)$ lies in ground-state  $|a)$  at  some   initial time
$t=0$.
Add   ``perturbation potential"  $W$   given by Eq.~(\ref{eq-W})  to    the ``unperturbed"   Hamiltonian  
defined by    the l.h.s. of
(\ref{eq-schroe3d}) where $i=a$ (ground state). Obviously, the effect of $W$
is simply to interchange the respective 
Coulomb interactions. 
Therefore the system becomes    time-dependent for $t>0$ . It is defined by
the    transition amplitude  
$\bigl(b|\Psi(r,t) \bigr)=\bigl( b | K_W(t,0) |a  \bigr)$   from ground state $|a)$  at initial time $t=0$
to excited eigenstate $|b)$  at some later time $t$ .
\begin{figure}[htbq]
 \includegraphics[width=0.60\textwidth,angle=0,bb=90 240  500 600,clip]{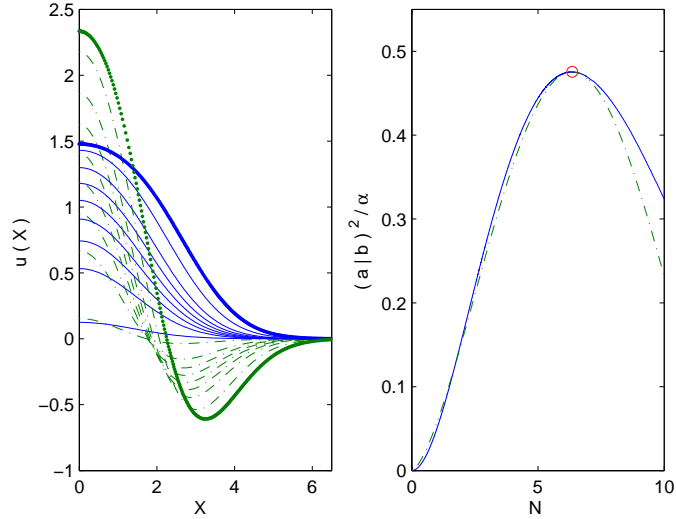}
      \caption{
\underbar{Left:} Several typical  dimensionless 
profiles $u(X)\propto \sqrt{{\cal N}}$  for the nonlinear ground-state   $|a)=|3/2)$
(continuous) and the next $s$ excited state
$|b)=|7/2)$ (broken-dotted)  when
${\cal N}=0.1,\, 0.4,\, 0.8,\, 1.3,\, 1.9,\, 2.7,\, 3.9,\, 6.5$ and $10$ 
(in bold). \underbar{Right:} The corresponding  square eigenstate overlap    
  (\ref{eq-normalisation}-\ref{eq-EqualN}) 
  as a function of  
${\cal N}$    displayed (for the sake of convenience) 
 in units of $\alpha=1/137.036$.
 The circle indicates the maximum   
$(a|b)_{max}^2 =3.4698911...\,10^{-3} \sim  0.4755\,\alpha$  occuring
at ${\cal N}={\cal N}_{max}=6.3542$. 
The broken-dotted line displays the  analytical approximation
$(a|b)^2 \sim 0.4755 \sin^{3/2}  [ (\pi/2) ({\cal N}/{\cal N}_{max}) ]$ for
 $0<{\cal N}\leq{\cal N}_{max}$.
}
\label{fig1}
\end{figure}
 The kernel
 $K_W(t,0) $  
for this  transition  is given by
  the implicit  integral equation:
\begin{equation}
\label{eq-kernel}
K_W(t,0)=K_{0}(t,0)-\frac{i}{\hbar}W\int_{0}^{t}  K_{ 0}(t,t')  K_W(t',0) dt',
\end{equation}
where the unperturbed kernel $K_{W\equiv 0}$, labelled $K_{0}$, becomes
 $K_{  0}(t,0)=\exp[-(i/\hbar)\mu_a t]$  and
$K_{  0}(t,t')=\exp[-(i/\hbar)\mu_b (t-t')]$ \cite{Feynman19:65} \cite{Cohen88:04}.
 The   lowest-order  transition amplitude in $W$,
defined by $K_W(t',0) \sim K_{  0}(t',0) $ in  integral  (\ref{eq-kernel}),
yields:
\begin{equation}
\label{eq-CsteMouvt}
\bigl( b | K_W(t,0) |a  \bigr)=e^{\displaystyle -\frac{i}{\hbar}\mu_a t }(b|a)-\frac{i}{\hbar}W_{ab}\,
e^{\displaystyle -\frac{i}{\hbar} \mu_b t }
\int_{0}^{t} e^{\displaystyle \frac{i}{\hbar}(\mu_b-\mu_a)t' } dt' + o[(W^2)_{ab}] + ... ,
\end{equation}
where $(b|a)=(a|b)$,  $W_{ab}=(\mu_b-\mu_a)(a|b)$ in accordance with Eq.~(\ref{eq-theoremW})
and $(W^2)_{ab}=(a|W^2|b)=(b|W^2|a)$.
Therefore:
\begin{equation}
\label{eq-goldenRexact}
\Bigl(b|\Psi(r,t) \Bigr)= \Bigl( b | K_W(t,0) |a  \Bigr)= (a|b)\exp \Bigl[ -i\frac{\mu_b}{\hbar} t\Bigr]    + o[(W^2)_{ab}] + ... ,
\end{equation}
and  the
   transition probability  from    
initial state $|a)$ to final state $|b)\neq |a)$ as a result of
     eigenstate-overlap  (\ref{eq-theoremW} - \ref{eq-W})
  becomes  time-independent
at  the lowest-order in $(a|b)\sim 6\,10^{-2}$:
\begin{equation}
\label{eq-goldenR}
\Bigl|\Bigl(b|\Psi(r,t) \Bigr)\Bigr|^2 \sim  (a|b)^2  = constant .
\end{equation}
Property   (\ref{eq-goldenR}) provides square eigenstate overlap $(a|b)^2$ with a clear
leading-order physical meaning in terms of a  time-independent  transition probability between $|a)$
and $|b)$. Equation (\ref{eq-goldenR})
   has been 
numerically checked up to $t=200/\omega$
\cite{Besse12:00}. It
 justifies  
 the further use of  time-independent
perturbation theory. 
Figure \ref{fig1} (right) displays $(a|b)^2$  as a function of   ${\cal N}$ 
for the two first $s$
(zero-angular-momentum)  nonlinear eigenstates $|a)=|3/2)$ and $|b)=|7/2)$ 
 illustrated by  Fig.  \ref{fig1} (left).
The half-integer   labels 
echoe their respective energy eigenvalues (in units of $\hbar \omega$) in the linear limit 
${\cal N}\rightarrow 0$. 
The maximum    eigenstate overlap    occurs  at  
${\cal N}={\cal N}_{max}=6.3542$  
and yields $(a|b)_{max}^2 = 3.4698911...\,10^{-3}$ (circle).

\section {Build-up of the soliton   Hamiltonian }

In order to construct  that soliton Hamiltonian ${\cal H}_{sol}$ which transforms  the   original       
\underbar{nonlinear}  system 
 Eqs ~(\ref{eq-schroeReduite} - \ref{eq-EqualN}) 
into an \underbar{equivalent linear}  invariant eigenproblem \cite{Bullough80}, \cite{Dodd82},\cite{Remoissenet99},
we follow the procedure explicitely given in Sections V.D.1 and V.D.2  
of  \cite {Cohen87:04} and consider the following     
 electromagnetic interaction Hamiltonian:
\begin{equation}
\label{eq-Hem}
 H_{em}=  e\int {\bold j} {\bold A}d^3 x=\Bigl(\frac{2\pi}{L}\Bigr)^{3/2}  e\sqrt{2\pi\hbar c} \int \Bigl[{\bold a}_{k'}\rho^*_{k'}+   
{\bold a}^+_{k'}\rho_{k'} \Bigr] \frac{d^3 k'}{\sqrt{k'}} .
\end{equation}
This Hamiltonian is in agreement with the choice of the classical Coulomb interaction potential
defined by Poisson Eq.\ (\ref{eq-poisson3d}) \cite{Lawrie90} \cite{Mandl59:59}.
In Eq.\ (\ref{eq-Hem}),  $c$ is the velocity of light,
 ${\bold j}$ is the  4-vector particle current density 
and   ${\bold A}$ its  related 4-potential operator while ${\bold a}^+_{k'}$ and ${\bold a}_{k'}$ are respectively
the  creation    and annihilation  operators  of  scalar photons 
with wave vector $k'$ and frequency $\omega_{k'}=c k'$.
Since our SP model  is  stationary at first order in accordance with Eq.\ (\ref{eq-goldenR}),  
   ${\bold j}$      reduces to its 4th static
charge-density component $\rho$ whose
   Fourier component is
$\rho_{k}$ \cite {Cohen87:04}. 
The choice of  $\rho$  quite naturally  occurs
  from 
  Coulomb potential (\ref{eq-W})
defining  eigenstate overlap (\ref{eq-theoremW}).
Indeed Poisson equation  $\nabla^2 W=-4\pi \rho$ yields:
\begin{equation}
\label{eq-rho}
 \rho(X)=e\bigl[ \Psi_b^2(X)-\Psi_a^2(X)\bigr]=\frac{e}{4\pi {\cal N} L^3}\bigl[ u_b^2(X)-u_a^2(X)\bigr] .
\end{equation}
Moreover, the soliton Hamiltonian ${\cal H}_{sol}$ 
should vanish in the linear limit ${\cal N}\rightarrow 0$ where there is neither Coulomb interaction
---hence no  photon exchange defined by electromagnetic Hamiltonian Eq.\ (\ref{eq-Hem})--- nor eigenstate overlap
Eqs\ (\ref{eq-theoremW} -\ref{eq-W}). 
Consequently we assume: 
\begin{equation}
\label{eq-HemNORMAL}
 {\cal H}_{sol}=f({\cal N}) H_{em} ,
\end{equation}
where   $f(0)=0$. 

As a result, soliton Hamiltonian ${\cal H}_{sol}$ defined by Eqs~(\ref{eq-Hem} - \ref{eq-HemNORMAL})
is indeed obtained 
from the   structural   properties of the   original       nonlinear  SP system 
through both its characteristic nonlinear parameter ${\cal N}$ and
 its square eigenstates $\Psi^2_{a,b}$ (or $u^2_{a,b}$ in their  dimensionless form).
It defines the ``nonlinear radiation perturbation" of ground state $|a)$ in the sense that
 it  takes into account both the electron-electron electrostatic interaction energy  in $|a)$ defined by Eq.\ (\ref{eq-Hem}) and
the  scaling  of this energy  by
eigenstate overlap  Eqs~(\ref{eq-theoremW} - \ref{eq-W}) (or equivalently by nonlinearity ${\cal N}$).
For the sake of simplicity, we start our search of function $f$
with the  easiest choice:
\begin{equation}
\label{ef2N}
f({\cal N}) \equiv {\cal N}.
\end{equation}
We shall see that this choice  is  indeed  acceptable in order to demonstrate the soliton property, but  for
that  very peculiar value of the harmonic electron confinement $\omega_{max}$ which corresponds to
${\cal N}={\cal N}_{max}$, i.e.  to maximum eigenstate
overlap.  We believe that the soliton property for any other value
$\omega$ of the electron confinement (i.e.  for any other value of ${\cal N}$) could indeed be demonstrated
through an appropriate choice of function $f({\cal N}) $ such that
$f(0) =0$ but this conjecture lies out of the scope of the present work.

The eigenstates of above-defined ${\cal H}_{sol}$ are obviously the QED photon-exchange  eigenstates of $H_{em}$.
Therefore the
 first-order amplitude  (in   ${\cal H}_{sol}$)  to create from   photon-vacuum
ground-state   $|a;0\rangle$
a photon  of energy $\hbar \omega_k=\hbar c k$   is
defined by Eqs~(\ref{eq-Hem} - \ref{ef2N}) and reads
    ${\cal A}_k={\langle k;a | {\cal H}_{sol}|a;0 \rangle}/(-\hbar\omega_k)$
in agreement with   time-independent perturbation theory \cite{Landau58}.
Hence:
\begin{equation}
\label{eq-PhotComponent}
{\cal A}_k=  
 \frac{\langle k;a |{\cal H}_{sol}|a;0 \rangle}{-\hbar c k} 
=- \Bigl(\frac{2\pi}{L}\Bigr)^{3/2}{\cal N} \sqrt{2\pi\alpha}\frac{n_k}{k^{3/2}} ,
\end{equation}
where $n_k$ 
is the  Fourier component  of
particle density $n=\rho/e$ given by (\ref{eq-rho}).
Amplitude (\ref{eq-PhotComponent}) displays, 
as expected, QED's single-photon leading-order    amplitude per particle $\propto \sqrt{\alpha}$ 
 \cite{Cohen88:04}
\cite{Mandl59:59}   \cite{Landau4:89})
where
$\alpha=e^2/(\hbar c)$ is the fine-structure constant.
Defining   $\kappa =L k$ and calling $G_{a,b}(\kappa)$ the 
dimensionless isotropic  radial
Fourier integral
\begin{equation}
\label{eq-g_k}
G_{a,b}(\kappa)=\int_0^{\infty} \bigl[ u_b^2(X)-u_a^2(X)\bigr]
\frac{\sin\kappa X}{\kappa X} X^2 dX ,
\end{equation}
we have $n_k=n({\kappa})=G_{a,b}(\kappa)/{\cal N} \sqrt{8\pi^3}$ and therefore:
\begin{equation}
\label{eq-Akappa}
{\cal A}_{\kappa}= - \sqrt{2\pi\alpha}\,\frac{G_{a,b}(\kappa)}{\kappa^{3/2}}.
\end{equation}

\section {soliton equivalence at maximum eigenstate overlap}

Let us now show  that 
 Hamiltonian  ${\cal H}_{sol}$
defined by use of Eqs~(\ref{eq-Hem} - \ref{ef2N})
 transforms  indeed the   original       nonlinear  system 
 Eqs ~(\ref{eq-schroeReduite} - \ref{eq-EqualN})
into the linear    photon-exchange eigenproblem
Eqs~(\ref{eq-Hem} - \ref{eq-Akappa})
for ${\cal N}={\cal N}_{max}$. This means that any physical  property
of  system 
 Eqs ~(\ref{eq-schroeReduite} - \ref{eq-EqualN}) at ${\cal N}={\cal N}_{max}$
---a priori non-observable due to   
non-orthogonality of the corresponding nonlinear eigenstates \cite{Huttner96:54}---
becomes actually
measurable 
since it can be recovered by use of the QED-like
linear eigenproblem Eqs~(\ref{eq-PhotComponent} - \ref{eq-Akappa}).
Specifically, the transition probability defined by
Eq.\ (\ref{eq-goldenR}) at maximum eigenstate overlap, as well as
the corresponding energy gap between nonlinear excited eigenstate   $|b)$ and
 ground state $|a)$
defined by Eqs~(\ref{eq-mu} - \ref{eq-greenfct}), should be recovered 
by use of Eqs~(\ref{eq-PhotComponent} - \ref{eq-Akappa}). We shall show
that this is indeed the case within the present first-order description yielding
a $\sim 1\%$ error.

\begin{figure}[htbq]
 \includegraphics[width=0.60\textwidth,angle=0,bb=90 240  500 600,clip]{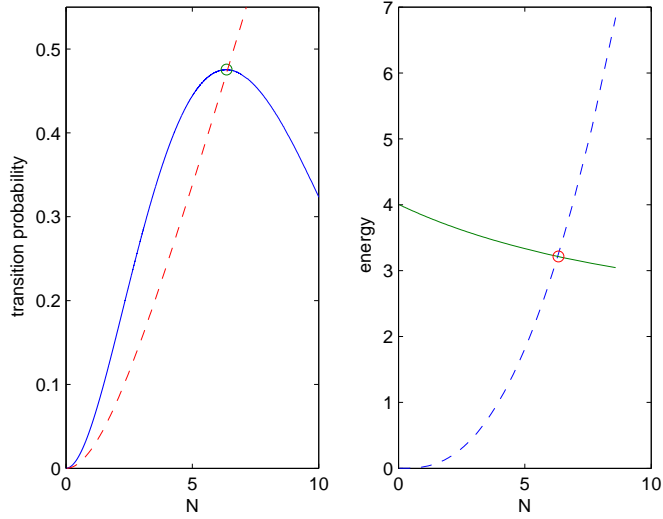}
      \caption{ SP (continuous line) vs QED-like soliton (broken line). \underbar{Left:} 
Photon-exchange   probability
${\cal P}$ defined by (\ref{eq-Pqed}) superimposed on the  eigenstate overlap    $(a|b)^2$ displayed 
  by the 
r.h.s. of
Fig. \ref{fig1}. 
At the  eigenstate overlap maximum located at ${\cal N}={\cal N}_{max}=6.3542$
and $(a|b)_{max}^2=3.4698911...\,10^{-3} $
(circle),  ${\cal P}=3.440\,10^{-3}$.
 \underbar{Right}:
 Quantum expectation value  ${\tilde {\cal E}}$ defined by (\ref{eq-Eqed})
superimposed on SP energy gap $\Delta  {\tilde E}={\tilde E_{7/2}}-{\tilde E_{3/2}}$  defined by (\ref{eq-Erond}).
The circle indicates
${\Delta  \tilde E}= 3.2145$   
at  maximum eigenstate overlap 
${\cal N}_{max}=6.3542$, to be compared with 
 ${\tilde {\cal E}}=3.2504$ for the same nonlinearity.
}
\label{fig2}
\end{figure}

The probability ${\cal P}$    for all possible transitions 
from SP ground state $|a)$ to the continuum of
virtual-photon energy states $\hbar\omega_k=\hbar c k = \hbar c \kappa/L$  above $|a)$
becomes:
\begin{equation}
\label{eq-Pqed}
{\cal P}=\sum_k |{\cal A}_k|^2 = \Bigl(\frac{L}{2\pi}\Bigr)^{3} \int_0^{\infty} |{\cal A}_k|^2 4\pi k^2 dk 
=\frac{\alpha}{\pi} \int_0^{\infty}G_{a,b}^2(\kappa)\frac{d\kappa}{\kappa}.
\end{equation}
Due to Eq.~(\ref{eq-EqualN})   which yields $G_{a,b}(0)=0$,
 there is no  $\kappa^{-1}$  divergence in Eq.~(\ref{eq-Pqed}), contrary to the
  well-known  logarithmic divergences in standard  QED
\cite{Landau4:89}.
Numerically calculating 
    integrals (\ref{eq-g_k})
and  (\ref{eq-Pqed})
for   values   $0<{\cal N}\leq 10$ yields the broken line ${\cal P}({\cal N})$ in 
Fig. \ref{fig2} (left). 
At ${\cal N}={\cal N}_{max}=6.3542$,   we obtain  ${\cal P}=3.440\,10^{-3}={\cal P}_{max}$
 which agrees with
  $\Pi_{max}=(a|b)_{max}^2=3.4698911...\,10^{-3}$ (circle)  within   the actual 
  $\sim 1\%$  error.  

Let us further show that the two   
probabilities  ${\cal P}_{max}$ and $ \Pi_{max}$  do  indeed define  the   same
transition of the electron pair  from  nonlinear
ground state  $|a)$ to its next nonlinear excited state
$|b)$ (remind:  both electrons  are  assumed to be in the same ``$s$"
orbital state).
 Once the nonlinear eigenvalue ${\tilde \mu}_i$ ($i=a,\,b$)
is   obtained by Eq.~(\ref{eq-mu})
for a given value ${\cal N}$,
the  quantum
expectation value $e \langle {\tilde \Phi}_i\rangle={\tilde \mu}_i- \langle C_i\rangle$ in eigenstate $|i)$
yields
the energy   per particle  \cite{Reinisch07:120402}
 ${\tilde \epsilon_i}=  {\tilde \mu}_i-\frac{1}{2}e \langle {\tilde \Phi}_i\rangle=
\frac{1}{2}({\tilde \mu}_i+\langle C_i\rangle )$. Therefore the \underbar{total} energy  for the electron pair 
in   $|i)$
 simply reads
\begin{equation}
\label{eq-Erond}
 {\tilde E_i}=2 {\tilde \epsilon_i} ={\tilde \mu}_i+\langle C_i\rangle =  {\tilde \mu}_i  +  
\frac{1}{{\cal N}}\int_0^{\infty}u^2_i (X) C_i(X) X^2dX.
\end{equation}
When $0<{\cal N}\leq 10$, the     energy gap 
$\Delta {\tilde E}={\tilde E_{7/2}}-{\tilde E_{3/2}}$ 
corresponding to $a=3/2$ and $b=7/2$
is displayed by the continuous line in Fig. \ref{fig2} (right). 
The value
$\Delta {\tilde E}= 3.2145={\tilde\Delta}_{max}$  at  
${\cal N}={\cal N}_{max}=6.3542$   is indicated by the circle.
It must   be compared with  the quantum expectation value \cite {Cohen87:04}:
\begin{equation}
\label{eq-Eqed}
{\tilde {\cal E}}= \frac{1}{\hbar\omega} \sum_k 
 \frac{\langle0;a|{\cal H}_{sol}|a;k \rangle
\langle k;a|{\cal H}_{sol}|a;0 \rangle}{-\hbar c k} 
=
\frac{1}{\hbar\omega}\sum_k |{\cal A}_k|^2 \hbar c k = \frac{{\cal N}}{\pi}
 \int_0^{\infty}G_{a,b}^2(\kappa) d\kappa ,
 \end{equation}
given in units of $\hbar\omega$ and  where the negative sign in the denominator of the series in Eq.~(\ref{eq-Eqed}
is cancelled by the property that the matrix elements of the creation and annihilation operators
do have always opposite signs
\cite {Cohen87:04}
 while   Fourier component  $G_{a,b}(\kappa)$ is defined by Eq.~(\ref{eq-g_k}).
 The mean energy ${\tilde {\cal E}}$ of the photon continuum
above the ground state  is displayed in broken line for $0<{\cal N} < 10$ in  Fig. \ref{fig2} (right).
 At   ${\cal N}={\cal N}_{max}$, it yields ${\tilde {\cal E}}=3.2504={\tilde {\cal E}}_{max}$, to be compared with
SP's  energy  gap ${\tilde\Delta}_{max}= 3.2145$  (circle).
The fit of these two energy
values 
within the estimated  $\sim 1\,\%$  error, together with the similar
 fit of the two corresponding 
transition probabilities displayed in Fig. \ref{fig2} (left)
at  maximum eigenstate overlap
${\cal N}={\cal N}_{max}$, 
   justifies  
the soliton transformation  defined by Eqs~(\ref{eq-Hem} - \ref{eq-Akappa}).

Note that the agreement ${\cal P}_{max}\sim \Pi_{max}$  at ${\cal N}={\cal N}_{max}$ actually yields  
(within the $\sim 1\%$ error bar)
a nonlinear definition of $\alpha$ which   solely depends on  the 
nonlinear eigenstates $|a)$ and $|b)$ of the non-relativistic
   quantum-dot model defined by
Eqs~(\ref{eq-schroe3d} - \ref{eq-poisson3d}). Indeed Eq.~(\ref{eq-Pqed}) yields at ${\cal N}={\cal N}_{max}$:   
\begin{equation}
\label{eq-alpha}
\alpha= \pi \frac{ \langle a | b \rangle_{max}^2}{\Biggl[\displaystyle \int_0^{\infty}G_{a,b}^2(\kappa)\frac{d\kappa}{\kappa}\Biggr]_{{\cal N}_{max}}}
=7.36082\,10^{-3} = \frac{1}{135.85} \sim \frac{1}{137.0359...}.   
\end{equation}
\section {possible experimental test for nonlinear eigenstate overlap }

Far-infrared spectroscopy was
 from early on used to investigate the electronic
structure   of quantum dots of various types. Soon it was realized that,
due to the extended Kohn theorem   \cite{Kohn61:123},
  such  a dipole radiation
whose wavelength is much larger than the dot size
yields a pure center-of-mass motion of the ``frozen"
electrons, 
independent of their number   and of the nature
of their interaction, provided that
the electron trap is harmonic \cite{Reimann02:74}. 
Therefore one has to investigate quantum
dots by use of 
resonant Raman scattering  \cite{Steinebach99:59}
\cite{Steinebach00:61}
   in order to analyze internal single-electron excitations and their
collective modes with monopole, dipole, or quadrupole
symmetry ($M = 0,±1,±2$ where $M$ is the quantum
number for angular momentum).
In particular, the $M = 0$ monopole
collective oscillations are excitations that can be exclusively
described by internal relative coordinates (e.g.  the breathing mode
that determines through its frequency
several ground state properties of a parabolic quantum
dot   \cite{McDonald13:111}). A  numerical simulation
of the excitation of such   an internal breather  mode has recently been performed \cite{Gudmundsson14:526}:
the quantum dot is radiated by a short oscillatory THz pulse of frequency
$\omega_{rad}$ whose spatial extend is of the order of
the harmonic length $L$. By extending the duration of the pulse, we 
can select more specifically that   radiation  frequency $\omega_{rad}$
which is resonant (in units of trap harmonicity $\omega$) with   the SP energy gap 
$\Delta {\tilde E}={ \tilde E_{7/2}}-{ \tilde E_{3/2}}$ 
displayed as a  continuous line in 
 Fig. \ref{fig2} (right).  For the sake of simplicity, approximate this latter  
(in units of $\hbar\omega$)
 by:
\begin{equation}
\label{eq-Deltalin}
\Delta {\tilde E}= 4 -0.124\, {\cal N}. 
\end{equation}
Now consider the particular   GaAs  isotropic
quantum-dot helium  where the effective  ``GaAs atomic unit"
$\eta=11.86$ meV is    deduced from   vacuum 
value 
$\eta_0=27.21$ eV   by use of  GaAs effective electron mass  $[m_e]_{GaAs}=0.067 \,m_e$   and 
dielectric constant   $12.4$. 
 Thence 
 the dependence of  resonant
$\omega_{rad}=\Delta {  E}/\hbar$  with respect to the trap harmonic frequency 
$\omega$ (in meV/$\hbar$) by use of
Eq.~(\ref{eq-Deltalin}):
\begin{equation}
\label{eq-omegaRESNT}
\hbar\omega_{rad}\Bigr|_{\Delta {\tilde E}}  \sim  \hbar\omega \Bigl(4-\frac{0.604}{\sqrt{\hbar\omega}}   
        \Bigr) \, \mathrm{meV},
 \end{equation}
since Eq.~(\ref{eq-N})  yields:
\begin{equation}
\label{eq-N3d}
{\cal N}=\sqrt{\frac{2 \eta}{\hbar\omega}} .
 \end{equation}
At  maximum eigenstate overlap ${\cal N}={\cal N}_{max}=6.3542$,  the corresponding
 confinement  frequency
$\omega= \omega_{max}$ becomes:
\begin{equation}
\label{eq-omegaMAX}
 \hbar\omega_{max}=\frac{2 \eta}{{\cal N}_{max}^2}=0.5873 \, \mathrm{meV} ,
 \end{equation}
Therefore a   GaAs  isotropic
quantum-dot helium  
initially at rest  in its ground state $|3/2)$ 
and then radiated by  an  external    pulse whose   frequency is 
tuned according to  Eq.\ (\ref{eq-omegaRESNT}) should display a
\underbar{minimum energy absorption} ---i.e.\ a minimum heating--- when   
its confinement frequency $\omega$
is varied
about  frequency (\ref{eq-omegaMAX}) because  the population of its excited eigenstate $|7/2)$ is maximum there
in agreement with Eq.~(\ref{eq-goldenR}). 

\section {conclusion }

By building up the specific    Hamiltonian   ${\cal H}_{sol}$ from both
the original   eigenstate properties of the two-electron
isotropic mean-field nonlinear SP model and the   single-photon 
electromagnetic  Hamiltonian,
we have shown a fundamental soliton equivalence between
the former 
\underbar{nonlinear} differential description and the latter   
\underbar{linear} QED-like    one.   
It is defined  by
the necessary  self-consistency  between   the classical  interacting
Coulomb potential that is used  in 
the  nonlinear  SP model  on the one hand,  and the corresponding selection 
of the sole single-photon  leading term in the   QED   series 
on the other hand. Hence the unavoidable   $\sim 1 \%$  error   of  the present model.
This soliton equivalence 
---which only  occurs    at maximum
eigenstate overlap in the present state of  theory--- 
allows   to translate the   nonlinear
(and hence non-observable)
quantum properties of the system into linear (and hence observable) ones.
Consequently  we suggest  modulated      
heating  of   radiated isotropic quantum-dot helium
by experimentally varying  its
 harmonic confinement and, thus, the population difference induced  by
       eigenstate overlap between  its two lowest-energy nonlinear eigenstates.  Finally we point out 
an intriguing nonlinear definition of the fine-structure constant
solely in terms of the non-relativistic SP eigenstates.

\begin{acknowledgments}
\noindent The author enjoyed        fruitful criticisms and/or 
 comments  
from    J.\ M.\ Raimond and  D.\ Est{\`e}ve. He   thanks
 B.\ Chauvineau,   P.\ DeLaverny, V.\ Gudmundsson  and, the last but not the least,  X.\ Waintal for 
 their   valuable  and encouraging help. 
He  feels deeply  grateful to P.\ Y.\ Longaretti and to the late P.\ Valiron for basic discussions 
at the onset of this research. Finally, he    acknowledges the hospitality
 provided by the University of Iceland and its
 Research
Fund.
\end{acknowledgments}
%


%

\end{document}